\documentclass[12pt]{article}

\title{Critical velocities $c/\sqrt 3$  and $c/\sqrt 2$ in general theory of relativity}
\author{S.I. Blinnikov, L.B. Okun, M.I. Vysotsky \\
ITEP, Moscow 117218, Russia}
\date{}
\begin{document}
\maketitle

\begin{abstract}

We consider a few thought experiments of radial motion of massive
particles in the gravitational fields outside and inside various
celestial bodies: Earth, Sun, black hole. All other interactions
except gravity are disregarded. For the outside motion there
exists a critical value of coordinate velocity ${\rm v}_c =
c/\sqrt 3$: particles with ${\rm v} < {\rm v}_c$ are accelerated
by the field, like Newtonian apples, particles with ${\rm v}
> {\rm v}_c$ are decelerated like photons. Particles moving inside
a body with constant density have no critical velocity; they are
always accelerated. We consider also the motion of a ball inside a
tower, when it is thrown from the top (bottom) of the tower and
after classically bouncing at the bottom (top) comes back to the
original point. The total time of flight is the same in these two
cases if the initial proper velocity $v_0$ is equal to $c/\sqrt
2$.
\end{abstract}

\section{Introduction. Proper and coordinate velocities}

According to the theory of relativity a time interval depends both
on the velocity of clocks and on the gravitational potential.
There exist infinitely many different coordinates. In the majority
of the physical experiments the gravitational potential remains
constant in the laboratory and it is convenient to use the
so-called proper time $\tau$ which is determined by the clocks at
rest in the laboratory frame. Maybe this is the reason why
sometimes in the literature the proper time is called ``genuine'',
or ``physical''. A lightminded person would think that any other
time (the coordinate time) is not physical, and thus should not be
considered. As a result some people working in the General Theory
of Relativity (GTR) consider coordinate-dependent quantities as
nonphysical, so to say ``second-quality'' quantities.

However  the coordinate time $t$ is even more important for some
problems than the proper time $\tau$. A good example is a particle
moving in a static gravitational field, which does not depend on
the time coordinate $t$. The proper time should be used when
discussing experiments performed in one and the same gravitational
potential (say, on a given floor of a house). However to discuss
events happening at different floors observers should introduce
corrections which take into account the difference of the
gravitational potentials. The coordinate time plays important role
in the global positioning system (GPS), where atomic clocks are
running on several dozens of the Earth satellites, and in
metrology in general \cite{6',7'}. That is why the coordinate
quantities are also ``genuine'', or ``physical''.

The solution of the Einstein equations outside a spherically
symmetric massive body was given by Schwarzschild \cite{8} and can
be found in all textbooks on GTR (see, for example,
\cite{4,6,10,7}). Simultaneously  Schwarzschild found the metric
inside a spherically symmetric body of constant density \cite{9}.
This paper is less known. Below we will compare the behaviour of
coordinate velocities in the Schwarzschild metrics \cite{8} and
\cite{9}. Some other coordinate systems will be considered at the
end of the paper.

As found in papers \cite{1} - \cite{3}, in the gravitational field
of a spherically symmetric object (the Earth, the Sun, other
stars) there exists a critical value of the coordinate speed
${\rm v}_c = c/\sqrt 3$. Particles moving radially with this speed
do not accelerate or decelerate in the first order of the Newton
constant $G$ -- thus, they ``ignore'' gravity.

For ${\rm v} < {\rm v}_c$ a falling object accelerates (a famous
example is the Newton apple). For ${\rm v} > {\rm v}_c$ a falling
object decelerates (a well known example is the electromagnetic
wave, or photon). This deceleration is a reason for the radar echo
delay (\cite{4}, chapter 8, \S7) and the deviation of light by the
Sun (\cite{4}, chapter 8, \S5).

That is why an intermediate critical coordinate velocity ${\rm
v}_c$ should exist.

We will show that outside the Sun ${\rm v}_c = c/\sqrt 3$, while
for a particle moving inside the Sun such a critical velocity does
not exist: any particle (even photon) accelerates moving to the
Sun center.

To study gravitational effects for particles moving inside
celestial bodies one should select such particles the
non-gravitational interaction of which with normal matter can be
neglected. Neutrinos, neutralinos and mirror particles \cite{5}
moving inside the Sun and the Earth satisfy this criterium.
Another example -- photons, electrons and nucleons moving inside a
mirror star. Finally particles moving inside a mine on the Earth
also satisfy the above condition.

In section 2 we introduce the necessary notations and obtain one
of the main results: ${\rm v}_c = 1/\sqrt 3$ for a particle moving
in a weak gravitational field: $r > R \gg r_g$, where $r$ is the
particle coordinate, $R$ -- the star radius and $r_g$ -- the
gravitational radius of the star \cite{4} - \cite{7}. Here and in
what follows we put $c=1$, that is why:
\begin{equation}
r_g = 2 GM \;\; , \label{1}
\end{equation}
where $M$ is the mass of the star  and $G$ -- the gravitational
constant.

In section 3 we consider the motion outside a star with strong
gravity (neutron star, black hole).

In section 4 we consider the motion inside a star, $r < R$, and
demonstrate that both apples and photons are accelerated when
freely falling inside a star.

In section 5 the notion of a critical proper (not coordinate!)
velocity is illustrated by the following  thought experiment: a
massive ball is thrown from the top (bottom) of the tower and
after elastically bouncing at the bottom (top) comes back after
time $\tau_+$ ($\tau_-$). The initial proper velocities in both
cases are equal $v_0$. For $v_0 \ll 1$ we have $\tau_+ < \tau_-$,
while for $v_0 =1$ we have $\tau_+ > \tau_-$. We found that
$\tau_+ = \tau_-$ for $v_0 = v_c \equiv 1/\sqrt 2$.

In section 6 other coordinate systems are considered (harmonic and
isotropic) which differ from the Schwarzschild one for $r \sim
r_g$, but describe flat space for $r \gg r_g$. In weak
gravitational fields ($r/r_g \gg 1$) the critical coordinate
velocity in these coordinates is also $1/\sqrt 3$ (plus
corrections $\sim (r_g/r)^2 \ll1$). Concerning  $v_c = 1/\sqrt 2$
it is invariant under a change of coordinates.

\section{Derivation of $\mbox{\boldmath${\rm v}_c = 1/\sqrt 3$}$
for $\mbox{\boldmath$r > R$}$}

For a radial motion ($\dot\theta = \dot\varphi =0$) the interval
looks like \cite{8}:
\begin{equation}
ds^2 = g_{00}dt^2 - g_{rr} dr^2 \equiv d\tau^2 - dl^2 \;\; ,
\label{2}
\end{equation}
where
\begin{equation}
g_{00} = 1-\frac{r_g}{r} \;\; , \label{3}
\end{equation}
\begin{equation}
g_{rr} = \left(1-\frac{r_g}{r}\right)^{-1} = \frac{1}{g_{00}} \;\;
. \label{4}
\end{equation}

The so-called local (or proper) velocity $v$ and coordinate
velocity ${\rm v}$ are the main subject of this paper. The local
velocity is measured by a local observer with the help of a ruler
and a clock which he has in hands. Doing this measurement the
local observer ignores that time is contracted and radius is
stretched by gravity:
\begin{equation}
v = \frac{d l}{d\tau} \;\; , \label{5}
\end{equation}
where the local (proper) coordinates $l$ and $\tau$ are defined in
Eq. (\ref{2}).

The same observer will measure the coordinate velocity if he takes
into account that his rulers and clocks are influenced by gravity
and uses the coordinates $r$ and $t$:
\begin{equation}
{\rm v} = \frac{d r}{d t} \;\; , \label{6}
\end{equation}
\begin{equation}
{\rm v} = v\left(\frac{g_{00}}{g_{rr}}\right)^{1/2} \;\; ,
\label{7}
\end{equation}
\begin{equation}
{\rm v} = v g_{00} \;\; . \label{8}
\end{equation}
Here the last equation follows from (\ref{7}) and (\ref{4}). From
(\ref{3}) and (\ref{4}) it follows that at  $r = \infty$ ~~
$g_{00}(\infty) = g_{rr}(\infty) =1$, that is why the coordinate
velocity coincides with the velocity measured by an observer who
resides infinitely far from the source of gravity (star).

When a body moves radially from $r=a$ to $r=b$ with a coordinate
velocity ${\rm v}(r)$, the following coordinate time elapses:
\begin{equation}
t = \int\limits^b _a \frac{dr}{\rm v} \;\; . \label{9}
\end{equation}
That is why to calculate the echo delay for a radar located far
from the star the coordinate velocity ${\rm v}$ is relevant.

For a particle freely moving in a static gravitational field the
conserved energy can be introduced:
\begin{equation}
E = \frac{m\sqrt{g_{00}}}{\sqrt{1-v^2}} \;\; , \label{10}
\end{equation}
(see \cite{6}, eq.(88.9)). The law of energy conservation,
\begin{equation}
E(r = \infty) = E(r) \;\; , \label{11}
\end{equation}
allows to determine $v(r)$:
\begin{equation}
1 - v^2 = (1-{\rm v}_\infty ^2)g_{00} \;\; . \label{12}
\end{equation}
Thus for a falling massive particle the proper velocity $v$
increases, while it does not change for a photon: $v_\gamma =1$.
${\rm v}(r)$ has a more complicated behavior:
\begin{equation}
1 -\frac{{\rm v}^2}{g_{00}^2} = (1-{\rm v}_\infty^2)g_{00} \;\; .
\label{13}
\end{equation}

For $r > R \gg r_g$, when the gravitational field is weak, from
(\ref{13}) and (\ref{4}) we get:
\begin{equation}
{\rm v}^2 = {\rm v}_\infty^2 +\frac{r_g}{r}(1-3{\rm v}_\infty^2)
\;\; . \label{14}
\end{equation}
For ${\rm v}_\infty \ll 1$  a well-known law for the motion of
nonrelativistic particle in gravitational field follows:
\begin{equation}
{\rm v}^2 = {\rm v}_\infty^2 +\frac{2MG}{r} \;\; . \label{15}
\end{equation}

For ${\rm v}_\infty = {\rm v}_c = 1/\sqrt 3$ the coordinate
velocity does not depend on $r$: ${\rm v} = {\rm v}_\infty$.

For ${\rm v}_\infty > {\rm v}_c$ a freely falling particle  slows
down!

Concerning a nonradial motion one easily finds that a massive body
is deflected by the Sun at any speed, and the deflection angle
$\theta$ is larger than that of light $\theta_\gamma$:
\begin{equation}
\theta = \frac{1}{2}\theta_\gamma(1+{\rm v}_\infty^{-2}) \;\; ,
\label{16}
\end{equation}
where
\begin{equation}
\theta_\gamma = \frac{2 r_g}{R} \;\; . \label{17}
\end{equation}
Here $R$ is the minimum distance between photon, radiated by the
distant star, and the Sun (see \cite{7}, problem 15.9, Eq. 13).

\section{Variation of $\mbox{\boldmath${\rm v}$}$ outside a black hole}

For a strong gravitational field $r_g/r \sim 1$ we get:
\begin{equation}
1-\frac{{\rm v}^2}{(1-\frac{r_g}{r})^2} = (1-{\rm v}_\infty^2)
(1-\frac{r_g}{r}) \;\; , \label{18}
\end{equation}
\begin{equation}
{\rm v}^2 = [1-(1-{\rm
v}_\infty^2)(1-\frac{r_g}{r})](1-\frac{r_g}{r})^2 \;\; .
\label{19}
\end{equation}

For ${\rm v}_\infty = 1/\sqrt 3$
\begin{equation}
{\rm v}^2 = \frac{1}{3}(1+\frac{2r_g}{r})(1-\frac{r_g}{r})^2 \;\;
. \label{20}
\end{equation}
For large values of $r$ ($r_g/r \ll 1$) one can neglect
$(r_g/r)^2$ and $(r_g/r)^3$ terms, and from (\ref{18}), (\ref{20})
it follows, that ${\rm v}^2 = 1/3$ and does not depend on $r$. For
$r_g/r \sim 1$ the behaviour of ${\rm v}(r)$ changes; in
particular it follows from  (\ref{19}) that ${\rm v}(r_g) =0$ for
any initial value ${\rm v}_\infty$.

For ${\rm v}_\infty > 1/\sqrt 3$ the coordinate velocity
monotonically decreases when $r$ decreases. Thus even moderately
relativistic particles behave like a photon.

For ${\rm v}_\infty < 1/\sqrt 3$ a particle starts to accelerate
and reaches a maximum coordinate velocity:
\begin{equation}
{\rm v}_{max} = \frac{2}{3\sqrt 3} \cdot \frac{1}{(1-{\rm
v}_\infty^2)} \label{21}
\end{equation}
at
\begin{equation}
r_{max} \equiv r({\rm v}_{max}) = \frac{3(1-{\rm
v}_\infty^2)}{(1-3{\rm v}_\infty^2)}r_g \label{22}
\end{equation}
After this  it starts to slow down, like a photon. Let us note
that
\begin{equation}
v({\rm v}_{max}) = 1/\sqrt 3 \;\; . \label{23}
\end{equation}

Thus the proper velocity equals $1/\sqrt 3$ at zero coordinate
acceleration.

\section{Variation of $\mbox{\boldmath${\rm v}$}$ inside a star}

As already stated in the Introduction we ignore all interactions
except gravitation.

The metric inside a celestial body of constant density
(oversimplified model of a star) was found by Schwarzchield
\cite{9} (see \cite{4}, section 11, \S 6).

For a radial motion:
\begin{equation}
g_{00} = \frac{1}{4}\left[3\sqrt{1-\frac{r_g}{R}}
-\sqrt{1-\frac{r_g r^2}{R^3}}\right]^2 \;\; , \label{24}
\end{equation}
\begin{equation}
g_{rr} = \left(1-\frac{r_g r^2}{R^3}\right)^{-1} \;\; , \label{25}
\end{equation}
where $R$ is  the radius of the star. Now $g_{rr} \neq 1/g_{00}$
and while Eq.(\ref{7}) is still valid, Eq.(\ref{8}) is not
satisfied. For a weak field, when $r_g/R \ll 1$ (let us note that
the condition $r_g/R < 1$ is necessary for star stability),
\begin{equation}
g_{00} = 1-\frac{3}{2} \frac{r_g}{R} +\frac{1}{2} \frac{r_g
r^2}{R^3} \;\; , \label{26}
\end{equation}
\begin{eqnarray}
{\rm v} & = & v\left(\frac{g_{00}}{g_{rr}}\right)^{1/2} = v\left[
1-\frac{3}{2}\frac{r_g}{R} -\frac{1}{2}\frac{r_g
r^2}{R^3}\right]^{1/2} = \nonumber \\ & = &
v\left(1-\frac{3}{4}\frac{r_g}{R} -\frac{1}{4} \frac{r_g
r^2}{R^3}\right) \;\; . \label{27}
\end{eqnarray}

We observe that even photons start to accelerate when they cross
the surface of a mirror star. (At the boundary $r=R$ the
coordinate acceleration of photons changes sign).

\section{A thought experiment in a tower and the critical proper velocity}

It is clear that the existence of the critical velocity ${\rm
v}_c$ is not connected with the infinite distance between the
observer and the gravitating body. Let us imagine a tower at the
Earth surface and an experimentalist with a clock having a metal
ball and a metal mirror which reflects this ball elastically. A
series of experiments is performed. Each experiment consists of
two parts. In the first part the experimentalist throws the ball
from the top of the tower with initial proper speed  $v_0 =
dl^+/d\tau^+$. The ball is reflected by the steel mirror situated
at the bottom of the tower and bounces back to the top. The local
time of this flight $\tau^+$ is measured at the top of the tower.
In the second part the experimentalist throws a ball from the
bottom of the tower upwards with initial velocity $v_0 =
dl^-/d\tau^-$. The ball is reflected by the mirror situated at the
top of the tower and the time interval of the flight $\tau^-$ is
measured. For nonrelativistic velocities  $v_0 \ll c$ we evidently
get $\tau^+ < \tau^-$. For photons $v_0 =1$ and $\tau^+ > \tau^-$.
The times $\tau^+$ and $\tau^-$ are equal for $v_0 = v_c = 1/\sqrt
2$. Thus, for $1/\sqrt 2 < v_0 \leq 1$ the ball is quicker when
thrown from the bottom. This puzzling result originates from the
time delay of the clock in a gravitational field; the same reason,
according to which the photons look redshifted when they move from
the bottom of the tower to the top (the famous Pound-Rebka
experiment).

Deriving the value of the critical velocity $v_c = 1/\sqrt 2$ we
will see that its value is universal: it does not depend on the
value of the gravitational potential and is valid for strong
fields as well. The only condition is: $h/r \ll 1- r_g/r$, where
$h$ is the height of the tower. Thus the tower should not be too
high (and tidal forces should not destroy the tower, the ball and
the experimentalist).

Let us get formulas which illustrate our statements.

The local time interval for the observer at the top of the tower
is:
\begin{equation}
d\tau^{+^2}  =  g_{00}^+ dt^2 \equiv (1-\frac{r_g}{r^+})d t^2 \;\;
,  \label{28}
\end{equation}
where $r^+$ is the radial coordinate of the top of the tower in
the Schwarzschield coordinates. The time of flight of the ball
according to the clock of this observer is:
\begin{equation}
\tau^+ = 2\int\frac{dr}{dr/d\tau^+} =
\frac{2}{\sqrt{1-\frac{r_g}{r^+}}}\int\frac{dr}{\frac{1}{(1-\frac{r_g}{r^+})}
\frac{dr}{dt}} \;\; . \label{29}
\end{equation}
This result immediately follows from Eqs.(\ref{9}) and (\ref{28}).

To find the expression for $dr/dt$ let us write the expression for
the local speed (which is coordinate invariant):
\begin{equation}
v = \left(\frac{g_{rr}}{g_{00}}\right)^{1/2} \frac{dr}{dt} =
\frac{1}{1-\frac{r_g}{r}}\frac{dr}{dt} \;\; , \label{30}
\end{equation}
and apply the energy conservation equation:
\begin{equation}
\frac{1-v_0^2}{1-\frac{r_g}{r^+}} = \frac{1-v^2}{1-\frac{r_g}{r}}
\;\; , \label{31}
\end{equation}
where $v_0$ is the initial speed of the ball. From Eqs.(\ref{30})
and (\ref{31}) for the square of the denominator of the integral
in (\ref{29}) we get:
\begin{equation}
\frac{1}{(1-\frac{r_g}{r^+})^2} \left(\frac{dr}{dt}\right)^2 =
\left(\frac{1-\frac{r_g}{r}}{1-\frac{r_g}{r^+}}\right)^2
\left[1-\frac{1-\frac{r_g}{r}}{1-\frac{r_g}{r^+}}(1- v_0^2)\right]
\;\; . \label{32}
\end{equation}

Expressing $r = r^+ - \varepsilon$ (where $\varepsilon$ varies
from zero to $h$,  $h$ is the height of the tower in
Schwarzschield coordinates) and expanding (\ref{32}) with respect
to $\varepsilon$ we obtain:
\begin{equation}
\frac{1}{(1-\frac{r_g}{r^+})^2}\left(\frac{dr}{dt}\right)^2 =
v_0^2 +\frac{r_g}{r^{+^2}(1-\frac{r_g}{r^+})}
\varepsilon(1-3v_0^2) \;\; , \label{33}
\end{equation}
where the weakness of gravitational field is unnecessary and the
only requirement is: $h \ll r^+ - r_g$.

Substituting Eq.(\ref{33}) into Eq.(\ref{29}) we obtain:
\begin{equation}
\tau^+ = \frac{2}{\sqrt{1-\frac{r_g}{r^+}}}
\int\limits_0^h\frac{d\varepsilon}{v_0\left[1+\frac{r_g(1-3v_0^2)}{2r^{+^2}
(1-\frac{r_g}{r^+})v_0^2}\varepsilon\right]} \;\; , \label{34}
\end{equation}
where we suppose that $v_0^2 \gg h r_g/r^{+^2} = 2gh$.

The same derivation applied for observer at the bottom of the
tower leads to the following expression for the time of flight of
the ball thrown by him:
\begin{equation}
\tau^- = \frac{2}{\sqrt{1-\frac{r_g}{r^-}}}
\int\limits_0^h\frac{d\varepsilon}{v_0\left[1-\frac{r_g(1-3v_0^2)}{2r^{-^2}
(1-\frac{r_g}{r^-})v_0^2}\varepsilon\right]} \;\; , \label{35}
\end{equation}
where $r^- = r^+ -h$.  For $v_0 = {\rm v}_c = 1/\sqrt 3$ integrals
in Eqs.(\ref{34}) and (\ref{35}) coincide, though $\tau^+$ will be
still smaller than $\tau^-$ because of the difference of the
factors which multiply integrals. To find the critical velocity
let us expand the expressions in eqs.(\ref{34}) and (\ref{35})
with respect to $\varepsilon$ and calculate the integrals:
\begin{equation}
\tau^+ = \frac{2h}{v_0\sqrt{1-\frac{r_g}{r^+}}} -
\frac{h^2}{v_0\sqrt{1-\frac{r_g}{r^+}}}
\frac{r_g(1-3v_0^2)}{2r^{+^2} (1-\frac{r_g}{r^+})v_0^2} \;\; ,
\label{36}
\end{equation}
$$ \tau^- = \frac{2h}{v_0\sqrt{1-\frac{r_g}{r^-}}} +
\frac{h^2}{v_0\sqrt{1-\frac{r_g}{r^-}}}
\frac{r_g(1-3v_0^2)}{2r^{-^2} (1-\frac{r_g}{r^-})v_0^2} \;\; . $$

Now we easily observe that these time intervals coincide for
$v_0\equiv v_c = 1/\sqrt 2$:
\begin{equation}
\tau^+ = \tau^- = \frac{2h}{v_c\sqrt{1-\frac{2r_g}{r^+ + r^-}}}
\;\; . \label{37}
\end{equation}

Performing an analogous experiment in a mine, an experimentalist
will find that $v_c$ is absent -- see Section 4.

\section{Isotropic and harmonic coordinates}

In this section we will demonstrate that the critical velocity
${\rm v}_c$ equals $1/\sqrt{3}$ in isotropic, harmonic and all
other asymptotically flat coordinates, related with Schwarzschield
coordinates by the following relations: $t^\prime = t$, $r^\prime
= r + {\rm const} + O(r_g^2/r^2)$.

According to well-known formulas  for the interval in isotropic
coordinates we have  (\cite{4}, chapter 8, \S\S 1-3) $$
ds^2=\frac{(1-MG/2\rho)^2}{(1+MG/2\rho)^2}dt^2
-\left(1+\frac{MG}{2\rho}\right)^4
(d\rho^2+\rho^2d\theta^2+\rho^2\sin^2\theta d\phi^2) \; . $$ Thus
in a weak field at  $\rho \to \infty$ and for a radial motion
($d\theta=0, \; d\phi=0$) we have: $$
ds^2=(1-2GM/\rho)dt^2-(1+2GM/\rho)d\rho^2 \; . $$

In harmonic coordinates: $$
ds^2=\left(\frac{1-MG/R}{1+MG/R}\right)dt^2
-\left(1+\frac{MG}{R}\right)^2d\mathbf{X}^2
-\left(\frac{1+MG/R}{1-MG/R}\right)\frac{M^2G^2}{R^4}(\mathbf{X}\cdot
d\mathbf{X})^2  \; . $$ where $$ X_1 = R\sin\theta\cos\phi, \quad
X_2 = R\sin\theta\sin\phi, \quad  X_3 = R\cos\theta \; , $$ and
$R^2 \equiv \mathbf{X}^2$. Thus at $d\theta=0, \; d\phi=0$: $$
ds^2=\left(\frac{1-MG/R}{1+MG/R}\right)dt^2
-\left[\left(1+\frac{MG}{R}\right)^2
+\left(\frac{1+MG/R}{1-MG/R}\right)\frac{M^2G^2}{R^2} \right] dR^2
\; . $$ In the limit $R \to \infty$ $$
 ds^2=(1-2GM/R)dt^2-(1+2GM/R)dR^2 \; .
$$ with the accuracy $O(r_g^2/R^2)$. Thus with this accuracy the
metric is the same in the Schwarzschield, isotropic and harmonic
coordinates.

Performing a derivation analogous to that of eqs. (18), (19) for
arbitrary coordinates $(r,t)$ in a spherically symmetric static
metric, where for a radial motion we can write $$
ds^2=g_{00}(r)dt^2-g_{rr}(r)dr^2 \; , $$ and generalizing this
derivation to the case when the initial proper velocity $v_0$ is
determined not at infinity, but at an arbitrary $r=r_0$, we get:
\begin{equation}
1-{\rm v}^2\frac{g_{rr}(r)}{g_{00}(r)} = (1-v_0^2)
\frac{g_{00}(r)}{g_{00}(r_0)} \; ,
\end{equation}
and
\begin{equation}
{\rm v}^2 = \left[1-(1-v_0^2)
\frac{g_{00}(r)}{g_{00}(r_0)}\right]\frac{g_{00}(r)}{g_{rr}(r)} \;
.
\end{equation}

In weak fields:

$$ g_{rr}(r)=\frac{1}{g_{00}(r)}+O\left(\frac{r_g^2}{r^2}\right)
\; , $$ the coordinate velocity is:
\begin{equation}
{\rm v}^2 = \left[1-(1-v_0^2)
\frac{g_{00}(r)}{g_{00}(r_0)}\right]g^2_{00}(r) \; .
\end{equation}
One easily observes that the coordinate acceleration is zero when
the local velocity  $v_0 = 1/\sqrt 3$ in all coordinates
considered by us. For these coordinates at infinity  ${\rm v} =v$,
and ${\rm v}_c = 1/\sqrt 3$ at $r_0=\infty$, precisely as stated
in the beginning of this section.

\section{Conclusion and acknowledgements}

We demonstrated that in a spherically symmetric gravitational
field two critical velocities exist: proper $v_c = 1/\sqrt 2$ and
coordinate ${\rm v}_c = 1/\sqrt 3$. The first is by definition
invariant under change of coordinates; the second is with high
accuracy the same in the Schwarzschield, isotropic and harmonic
coordinates, which are widely used to synchronize clocks on
satellites, to analyze the double neutron stars motion and for
other relativistic astrophysical objects. We criticized the
terminology according to which the proper time is called
``physical'' or ``genuine''.

The existence of critical velocities $c/\sqrt 3$ and $c/\sqrt 2$
does not contradict the unique role played by light velocity $c$
in the relativistic domain. However the acquaintance with the
critical velocities in Schwarzschield coordinates will help the
reader to study better the general theory of relativity since
these coordinates are ``sensible, useful and frequently used''
(see \cite{10}, v.2, p.526).

We are grateful to S.~Deser and B.~Tekin who pointed papers
\cite{1} and \cite{2} to us (after we published the paper
\cite{3}).

We are grateful to the UFN anonymous referee of this paper. Due to
his efforts the paper was substantially expanded.

We are partially supported by FS NTP FYaF 40.052.1.1.1112. SB is
partly supported by RFBR grant 02-02-16500 and ILE in the Osaka
university. LO is partly supported by A. von Humboldt Award.

\end{document}